\documentclass[aip, twoside, reprint]{revtex4-1}

\usepackage{graphicx}
\usepackage[version=3]{mhchem}
\newcommand{\angstrom}{\mbox{\normalfont\AA}}

\begin{document}
	
	\title{The Breakdown of Mott Physics at \ce{VO2} Surfaces}
	
	\author{Matthew J. Wahila} \affiliation{Department of Physics, Applied Physics and Astronomy, Binghamton University, Binghamton, New York 13902, USA\looseness=-1}
	
	\author{Nicholas F. Quackenbush} \affiliation{Department of Physics, Applied Physics and Astronomy, Binghamton University, Binghamton, New York 13902, USA\looseness=-1}

	\author{Jerzy T. Sadowski}\affiliation{Center for Functional Nanomaterials, Brookhaven National Laboratory, Upton, New York 11973, USA\looseness=-1}
	
	\author{Jon-Olaf Krisponeit}\affiliation{Institute of Solid State Physics, University of Bremen, Otto-Hahn-Allee 1, 28359 Bremen, DE\looseness=-1}
	\author{Jan Ingo Flege}\affiliation{Institute of Solid State Physics, University of Bremen, Otto-Hahn-Allee 1, 28359 Bremen, DE\looseness=-1}\affiliation{Applied Physics and Semiconductor Spectroscopy, Brandenburg University of Technology Cottbus-Senftenberg,	Konrad-Zuse-Str. 1,	03046 Cottbus, DE\looseness=-1}
	
	\author{Richard Tran} \affiliation{Department of NanoEngineering, University of California San Diego, 9500 Gilman Drive 0448, La Jolla, California 92093, USA\looseness=-1}
	\author{Shyue Ping Ong} \affiliation{Department of NanoEngineering, University of California San Diego, 9500 Gilman Drive 0448, La Jolla, California 92093, USA\looseness=-1}
	
	\author{Christoph Schlueter}\affiliation{Diamond Light Source Ltd., Diamond House, Harwell Science and Innovation Campus, Didcot, Oxfordshire OX11 0DE, UK\looseness=-1}
	\author{Tien-Lin Lee}\affiliation{Diamond Light Source Ltd., Diamond House, Harwell Science and Innovation Campus, Didcot, Oxfordshire OX11 0DE, UK\looseness=-1}
	
	\author{Megan E. Holtz}
	\affiliation{School of Applied and Engineering Physics, Cornell University, Ithaca, NY 14853, USA\looseness=-1}
	\author{David A. Muller}\affiliation{School of Applied and Engineering Physics, Cornell University, Ithaca, NY 14853, USA\looseness=-1}\affiliation{Kavli Institute at Cornell for Nanoscale Science, Ithaca, New York 14853, USA\looseness=-1}
	
	\author{Hanjong Paik}\affiliation{Department of Materials Science and Engineering, Cornell University, Ithaca, New York 14853-1501, USA\looseness=-1}
	\author{Darrell G. Schlom}\affiliation{Department of Materials Science and Engineering, Cornell University, Ithaca, New York 14853-1501, USA\looseness=-1}\affiliation{Kavli Institute at Cornell for Nanoscale Science, Ithaca, New York 14853, USA\looseness=-1}
	
	\author{Wei-Cheng Lee} \affiliation{Department of Physics, Applied Physics and Astronomy, Binghamton University, Binghamton, New York 13902, USA\looseness=-1}
	
	\author{Louis F. J. Piper}\email{lpiper@binghamton.edu} \affiliation{Department of Physics, Applied Physics and Astronomy, Binghamton University, Binghamton, New York 13902, USA\looseness=-1} \affiliation{Materials Science {\&} Engineering, Binghamton University, Binghamton, New York 13902, USA\looseness=-1}

	\begin{abstract}
		Transition metal oxides such as vanadium dioxide (\ce{VO2}), niobium dioxide (\ce{NbO2}), and titanium sesquioxide (\ce{Ti2O3}) are known to undergo a temperature-dependent metal-insulator transition (MIT) in conjunction with a structural transition within their bulk. However, it is not typically discussed how breaking crystal symmetry via surface termination affects the complicated MIT physics. Using synchrotron-based x-ray spectroscopy, low energy electron diffraction (LEED), low energy electron microscopy (LEEM), transmission electron microscopy (TEM), and several other experimental techniques, we show that suppression of the bulk structural transition is a common feature at \ce{VO2} surfaces. Our density functional theory (DFT) calculations further suggest that this is due to inherent reconstructions necessary to stabilize the surface, which deviate the electronic structure away from the bulk d$^1$ configuration. Our findings have broader ramifications not only for the characterization of other "Mott-like" MITs, but also for any potential device applications of such materials.
	\end{abstract}

	\maketitle
	Bulk, unstrained vanadium dioxide (\ce{VO2}) undergoes a temperature-dependent metal-insulator transition (MIT) at 65~$^\circ$C accompanied by a crystal symmetry change between the high-temperature rutile and low-temperature M1 phases. After decades of investigation and debate,\cite{Kosuge1967,Eyert2002} \ce{VO2} is now generally understood to be a cooperative Mott-Peierls system with aspects of both playing an integral role in the MIT. However, the exact interplay between the electrons (Mott) and lattice (Peierls) is still being unraveled.\cite{Qazilbash2007,Quackenbush2016M2,Quackenbush2015,MukherjeePRB2016,Quackenbush2017,WCLee2019} For example, there is currently interest in determining if the electronic and structural phase transitions can be truly decoupled under specific circumstances, although studies reported so far have been plagued by issues such as poorly-defined sample quality, non-contemporaneous measurement of the structure and electronic properties, and other issues.\cite{Lee2018,Yang2016,Tao2012,Laverock2014,Laverock2018} Despite these lingering fundamental questions, the abrupt and large discontinuity in resistivity across the MIT has generated significant interest in \ce{VO2} for its' potential uses in novel switching devices.\cite{Yang2011, Zhou2015, Chen2015, Chen2019} 
	
	\ce{VO2} has shown promise for many applications, such as in ionic liquid gated, field-effect transistors that undergo bulk carrier delocalization instigated by charge accumulation at the surface.\cite{Nakano2012,Jeong2013} This and other \ce{VO2} devices mainly employ ultra-thin epitaxial films, as the MIT properties of these films can be fine-tuned for specific applications using the strain induced by epitaxial matching with the substrate.\cite{Muraoka2002,Fischer2020} Unfortunately, as the dimensions of these \ce{VO2} films decrease, the film surface/interface properties become increasingly important for device performance, while relatively little is known about the intrinsic nature of \ce{VO2} surfaces.\cite{Mellan2012}
	
	Here, we employ low energy electron diffraction (LEED) and microscopy (LEEM) to perform a temperature-dependent surface study on three orientations of high quality, epitaxial \ce{VO2} films; (100), (001), and (110). Multiple x-ray spectroscopy techniques, including x-ray absorption and x-ray photoelectron spectroscopies (XAS \& XPS), are used in conjunction to monitor the electronic structure and composition of the films throughout the study. None of our temperature-dependent measurements provide any evidence of a structural change at the surface corresponding to the bulk MIT. Mild annealing and/or electron beam exposure is found to induce some morphological changes, however, these are explained purely by general surface stability arguments. This study reveals and explains the existence of unavoidable surface layer reconstructions that do not directly participate in the bulk \ce{VO2} MIT, regardless of film orientation.
	
	\begin{figure*}[htb!]
		\includegraphics{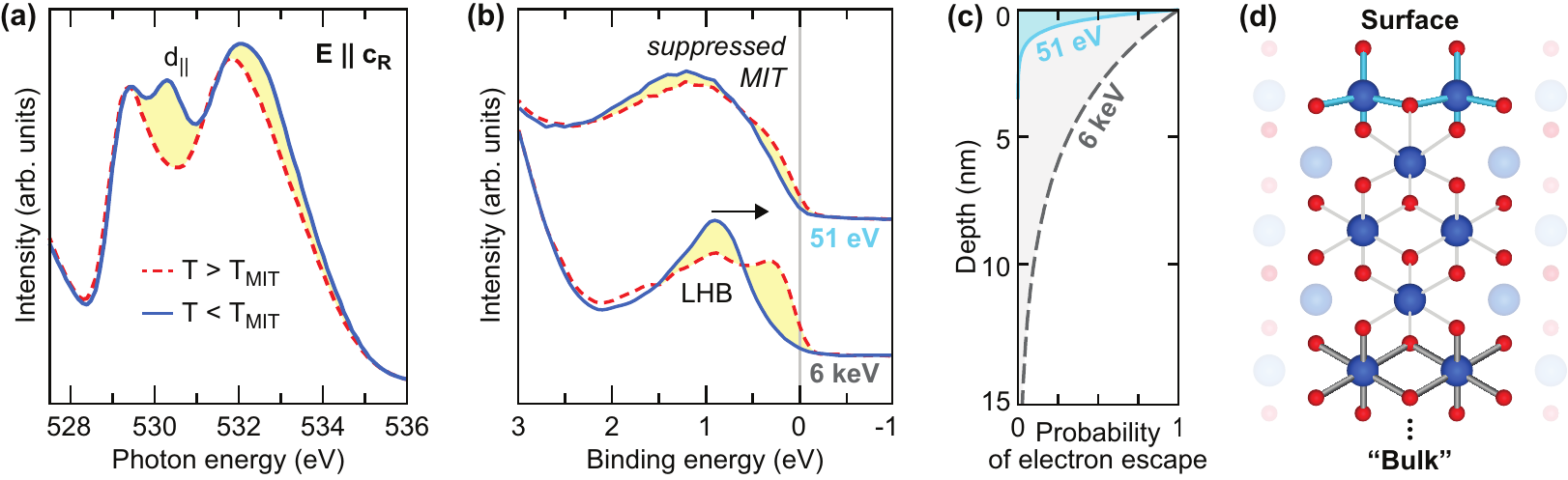}
		\caption{\label{VO2_100summary}  \textbf{(a)} Oxygen K-edge XAS and \textbf{(b)} valence band UPS/HAXPES of a \ce{VO2}/\ce{TiO2} (100) film measured above and below T$_{MIT}$. \textbf{(c)} Approximate photoelectron escape probabilities with depth below sample surface for the photon energies used in this experiment ($\sim$6~keV for HAXPES and 51~eV for UPS). \textbf{(d)} The most stable DFT calculated structure for a stoichiometric \ce{VO2}(100) slab terminating in vacuum.}
	\end{figure*}

	\section{Probing the MIT/SPT of the \ce{VO2} epilayer}

	The structural transition of a clean \ce{VO2} film can be investigated using oxygen K-edge XAS, as shown in Fig. \ref{VO2_100summary}a. Due to the strong O 2p - V 3d hybridization in \ce{VO2}, the O K-edge provides an indirect measure of the V-V dimer formation associated with the MIT. The presence of dimers in the insulating M1 phase is evidenced by the so-called d$_{\parallel}$ feature at $\sim$530.5~eV, located between the $\pi$* ($\sim$529.5~eV) and $\sigma$* ($\sim$532~eV) states.\cite{Quackenbush2015, GaloPRL} In \ce{VO2}(100), this d$_{\parallel}$ feature completely vanishes above the MIT temperature, indicating a transition from the M1 to rutile phase. This d$_{\parallel}$ feature can also provide insight into the relative level of electron correlation present in films of different orientations (see Supplementary Figure 1).
	
	Figure \ref{VO2_100summary}b shows complimentary UV and hard x-ray photoelectron spectroscopies (UPS \& HAXPES) of the topmost valence states of a \ce{VO2}(100) epitaxial thin film. The top valence states (0 - 2~eV) are mainly V 3d in character and lie just above the predominantly O 2p states that constitute the majority of the valence band. We observe clear differences in the V 3d states above and below the MIT temperature using HAXPES, with the density of states increasing at the Fermi level (E$_F$) in the rutile phase indicating an increased metallicity. In contrast, a much broader V 3d peak with a somewhat suppressed transition is observed using the more surface sensitive UPS.
	
	As shown in \ref{VO2_100summary}c, the probability of photoelectron escape with increasing depth inside the sample drops off much more quickly for UPS compared to HAXPES. In fact, the probing depth of UPS is only on the order of 1-2 nm, while that of HAXPES is closer to $\sim$15~nm. Therefore, the difference in the observed spectral weight transfer between UPS and HAXPES implies that the MIT does not occur as strongly (if at all) within the topmost atomic layers of the \ce{VO2}(100) film. 
	
	The DFT calculated most stable surface structure for a stoichiometric \ce{VO2}(100) slab in vacuum, shown in Figure \ref{VO2_100summary}d, supports the idea of a suppressed transition within the film surface layers. The break in crystal symmetry at the surface is predicted to result in a surface reconstruction imparting the topmost vanadium atoms with a different oxygen coordination than those in the bulk. This is in agreement with previous studies that report an ultra-thin V$^{5+}$ surface layer on \ce{VO2} films, even after \textit{in vacuo} cleaning and surface preparation.\cite{Quackenbush2015} 
	
	Furthermore, even for those vanadium atoms with the expected "bulk" coordination, some distortion of inter-atomic bond distances is predicted within the near surface layers of the slab. As both doping and strain are known to drastically alter the \ce{VO2} MIT characteristics by pushing the d-filling closer to or farther from the Mott criterion,\cite{WCLee2019} the observed alterations to the coordination and inter-atomic distances predicted at the terminated \ce{VO2} surface should also be expected to alter the MIT characteristics.
	
	\section{Suppression of the transition at the surface}
	
	\begin{figure*}[htb!]
		\includegraphics{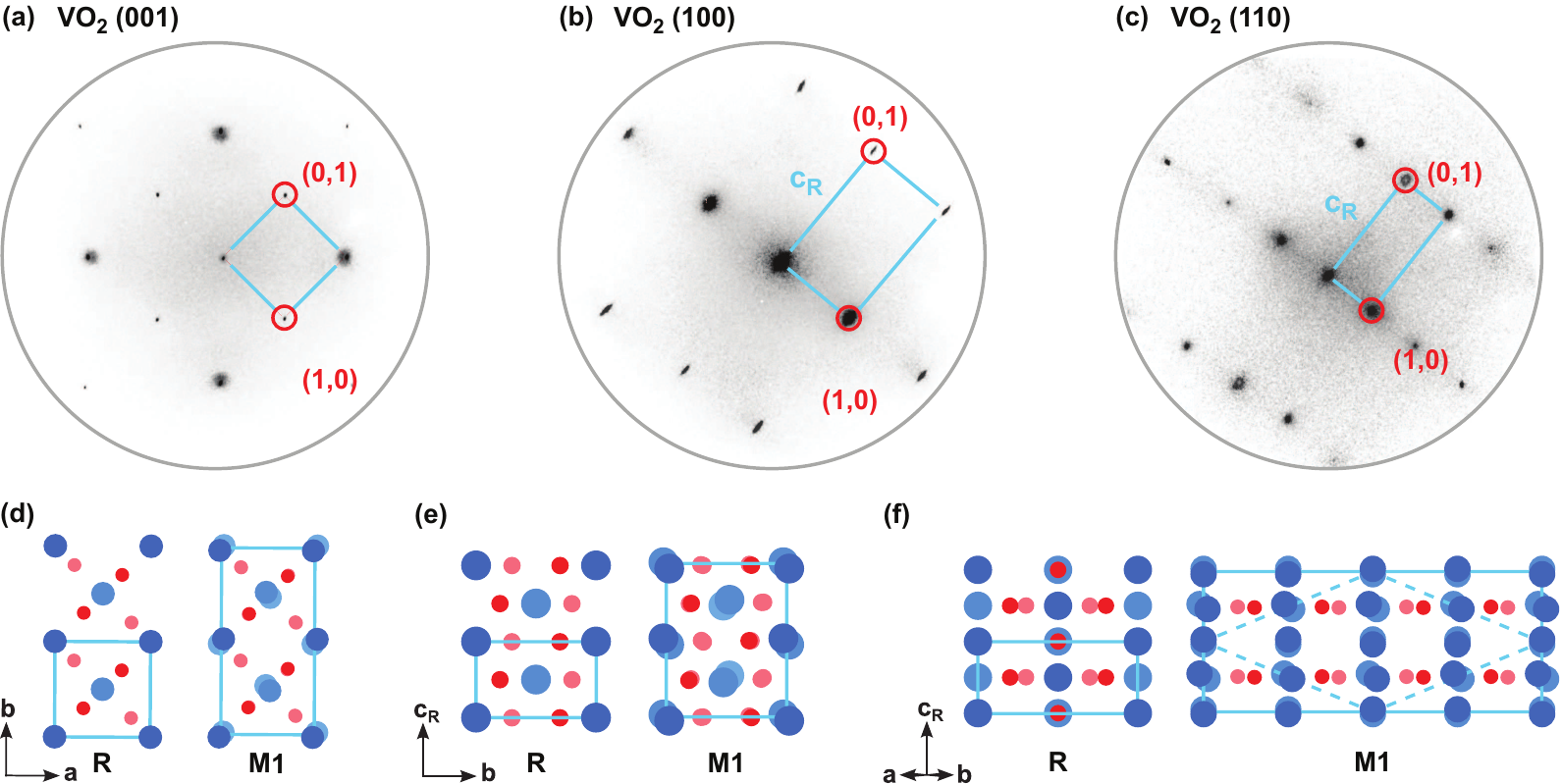}
		\caption{\label{VO2_LEEDsurfaces} Averaged LEED patterns for \textbf{(a)}~\ce{VO2}(001), \textbf{(b)}~\ce{VO2}(100), and \textbf{(c)}~\ce{VO2}(110) epitaxial thin films on \ce{TiO2} single crystal substrates, along with structural models of the expected rutile and equivalent M1 surfaces. Vanadium and oxygen atoms are represented by large blue and small red circles, respectively. Surface unit cells are indicated by light blue rectangles.}
	\end{figure*}
	
	Shown in Figure \ref{VO2_LEEDsurfaces}a, the LEED pattern from an as-loaded \ce{VO2}(001) surface displays square (p4mm) symmetry with lattice constants comparable to that of the corresponding rutile \ce{TiO2} ($1\times1$) surface. This is an averaged LEED pattern of those measured at 37, 47, 50, and 73~eV beam energy, due to the many spectral extinctions observed on this surface. Notably, a sharp (0,0) spectral spot is not observed until the highest beam voltage. We observed no changes to the LEED pattern upon cooling or heating over a temperature range of -30--130$^\circ$C. 
	
 	As epitaxial \ce{VO2}/\ce{TiO2}(001) films undergo their MIT near room temperature, any MIT-related structural changes should have been observed during these temperature-dependent experiments.\cite{WCLee2019,Quackenbush2016M2} Since the c$_R$ crystal axis is out-of-plane for the (001) orientation, the V-V dimerization in the M1 phase will occur perpendicular to the film surface, which might be expected to show no signature in the LEED. However, the concomitant zig-zagging of vanadium atoms in the M1 phase should result in a doubling of the periodicity along one direction at the surface if the film is terminated along a rutile (001) equivalent plane, as depicted by the structure diagram in Fig. \ref{VO2_LEEDsurfaces}d. This periodicity doubling should correspond to half-order spots in the LEED pattern, however, none were observed during the experiments.
 	
	Shown in Fig. \ref{VO2_LEEDsurfaces}b, the LEED pattern from an as-loaded \ce{VO2}(100) surface displays rectangular (p2mm) symmetry with lattice constants comparable to that of the corresponding rutile \ce{TiO2} ($1\times1$) surface. This is an averaged LEED pattern measured at 30 and 60~eV beam energy. This surface shows a (0,0) spot at most energies, and spectral extinctions occur most often in second order diffraction spots. Also, each diffraction spot on this surface is slightly elongated along the (0,1) direction. 

	In contrast to the (001) orientation, \ce{VO2}/\ce{TiO2}(100) is oriented such that the c$_R$ crystal axis is in-plane with the surface. Upon entering the low temperature phase, the lowering of symmetry due to V-V dimerization should result in a clear doubling of the unit cell along the c$_R$ axis, as depicted by the structure diagram in Fig. \ref{VO2_LEEDsurfaces}e, which would be evident in the LEED as half-order spots long the (0,1) direction.\cite{Cao2009} In addition, this in-plane c$_R$ axis is elongated by the epitaxial strain, which increases the MIT temperature and also introduces the possibility of other intermediate phases within the epilayer that may have signatures in the LEED pattern, such as the M2 structure.\cite{Quackenbush2016M2,Park2013} However, no intermediate spots were observed during the experiments.
	
	Although the \ce{VO2}(100) epilayer is known to be in the M1 phase at room temperature, the observed room temperature surface unit cell corresponds to a ($1\times1$) reconstruction of a bulk termination of the rutile unit cell. And this rutile-like reconstruction is found to be stable from room temperature up to at least 150~$^{\circ}$C. This pattern is notably different to the only other \ce{VO2}(100) surface reported in the literature, which indeed showed half-order spots, although they were observed along the direction corresponding to the rutile b axis.\cite{Goering1997}
	
	The \ce{VO2}(110) surface is similar to the \ce{VO2}(100) surface in that the c$_R$ axis is completely in-plane, however, the exposed surface for this orientation is a diagonal cut through the 3D unit cell. The LEED pattern for this as-loaded surface is shown in Fig. \ref{VO2_LEEDsurfaces}c. The pattern is similar to the bulk terminated ($1\times1$) rutile \ce{TiO2}(110) pattern, and corresponds to a ($1\times1$) rutile-like surface reconstruction of \ce{VO2}. 
	
	As depicted by the structure diagram in Fig. \ref{VO2_LEEDsurfaces}f, terminating the M1 structure along a rutile (110) equivalent plane would be expected to produce surface ordering with a greatly expanded unit cell. This should produce fractional spots along both the (0,1) and (1,0) directions in the LEED. However, similar to the other orientations, this surface is found to show no temperature dependence and remain rutile-like at all measured temperatures above and below the MIT of the epilayer.

	\section{Stability of different \ce{VO2} surfaces}
	
	\begin{figure*}[htb!]
		\includegraphics{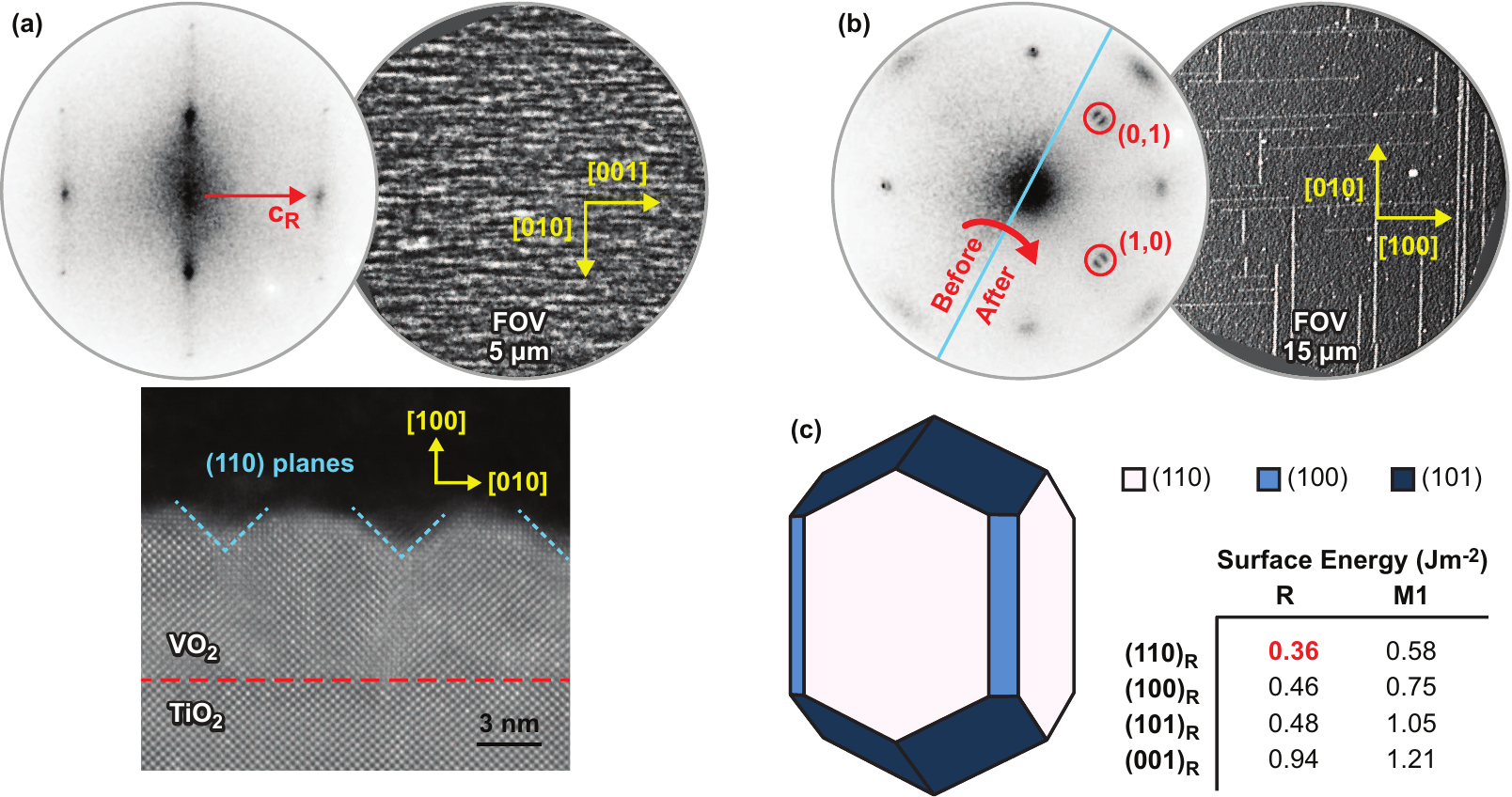}
		\caption{\label{VO2_TEM_LEEM} \textbf{(a)} LEED (43 eV), LEEM, and ADF STEM of a 7.5~nm \ce{VO2}/\ce{TiO2}(100) film after annealing showing the formation of (110) facets on the film surface. \textbf{(b)} Composite of LEED (25 eV) images taken before/after annealing a 10~nm \ce{VO2}/\ce{TiO2}(001) film and the corresponding LEEM showing the existence of micro-cracks after annealing. \textbf{(c)} Calculated Wulff shape of the rutile structure,  and surface energies for common surfaces of the \ce{VO2} rutile and monoclinic crystal structures. Terminations were non-polar to obtain the lowest surface energy.}
	\end{figure*}
	
	Shown in Figure \ref{VO2_TEM_LEEM}a, the \ce{VO2}(100) LEED pattern begins to display a pronounced elongation perpendicular to the c$_R$ direction following several anneal cycles (P$_{O_2}$ = $1.0\times10^{-6}$~Torr). The corresponding LEEM image shows the development of a highly directional, corrugated surface texture running perpendicular to the c$_R$ direction, consistent with the striations observed in the LEED. Cross-sectional scanning transmission electron microscopy (STEM) annular dark field (ADF) images of the \ce{VO2} specimen viewed along the [001] axis reveal nature of the surface texture. The anneal cycles induced the formation of (110) facets along the surface, parallel to the [001] direction. On this specimen, these facets have an average spacing of approximately 8.7~nm. Literature reports on \ce{TiO2}(100) indicate that it also has a tendency to reconstruct and form (110) micro-facets.\cite{Diebold2003, Hardman1993, Landree1998}
	
	Typically, changes in the observed \ce{VO2}(001) LEED pattern are very small after several anneal cycles. On average the background is reduced slightly, although the diffraction spots can also become less sharp with indications of some vague faceting along the rutile [110] directions.\cite{Falta2020} However, on occasion it is possible to observe another LEED pattern for the \ce{VO2}(001) surface wherein the first order spectral spots completely split. Shown in Figure \ref{VO2_TEM_LEEM}b, this new LEED pattern is associated with heating through the MIT, although it is completely irreversible once formed and therefore not associated with V-V dimerization. Corresponding LEEM displays long, straight streaks on the film surface interpreted as micro-cracks formed to relieve mechanical stresses within the film produced by the epitaxial strain. Previous reports indicate similar cracks tend to form along the [010] and [100] directions during the growth of much thicker films ($\sim$30 nm),\cite{Nagashima2006,Paik2015} while recent investigations by Krisponeit \textit{et al.} report the appearance of micro-cracks post-growth due to thermal cycling.\cite{Falta2020}
	
	DFT calculations of surface energies, given in Figure \ref{VO2_TEM_LEEM}c, provide some insight into the observed behavior of these \ce{VO2} films following annealing. All rutile \ce{VO2} surfaces are predicted to be slightly lower energy than their respective M1 counterparts, with the rutile (110) surface being the lowest energy of all. This suggests \ce{VO2} surfaces may reconstruct into a rutile pattern regardless of the bulk structure. In addition, \ce{VO2} would prefer to expose the (110) surface if possible, explaining the (110) micro-facets observed in (100) orientation films.

	\section{Oxygen-related reconstructions of the (110) surface}
	
	\begin{figure*}[htb!]
		\includegraphics{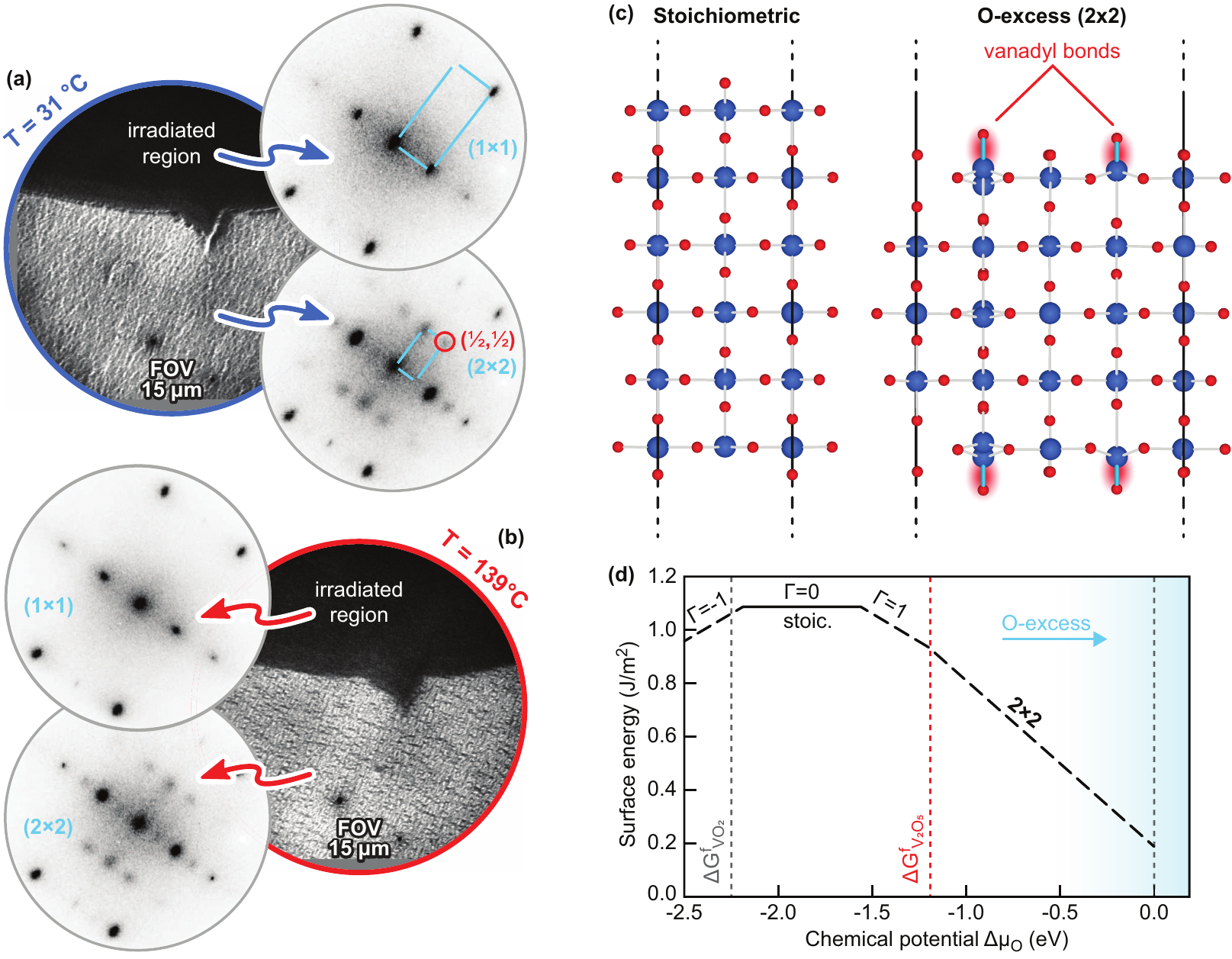}
		\caption{\label{VO2_110_damage} LEEM ($\sim$1.5 eV) and LEED (30 eV) of a \ce{VO2}/\ce{TiO2}(110) film comparing a region irradiated with the electron beam for several minutes to a non-irradiated region, both \textbf{(a)} below and \textbf{(b)} above the MIT. \textbf{(c)} DFT predicted stoichiometric surface and oxygen-rich (2$\times$2) surface reconstruction of a \ce{VO2}(110) slab. \textbf{(d)} DFT predicted surface energies of a \ce{VO2}(110) slab with oxygen chemical potential ($\Delta \mu_O$). $\Delta \mu_O =0$ eV is set to be the oxygen chemical potential of \ce{O2} gas. For context, we drew vertical dashed lines for the formation energy per oxygen atom of \ce{VO2} and \ce{V2O5} ($\Delta G^f_{\ce{VO2}}$ and $\Delta G^f_{\ce{V2O5}}$) and the enthalpy of reaction for \ce{V2O5} relative to \ce{VO2} ($\Delta H^f_{\ce{V2O5}}$).}
	\end{figure*}
	
	After performing anneals on \ce{VO2}(110), some slight surface striations can appear in LEEM while the LEED spots become more diffuse and some higher-order spots become visible, similar to the (100) surface behavior. The post-anneal texture on this surface, however, appears different to that of the (100) surface and is more likely due to the formation of step terraces, as this is known to occur on \ce{TiO2}(110) surfaces.\cite{Diebold2003}
	
	Shown in Figure \ref{VO2_110_damage}a and b, it is possible to have a ($2\times2$) reconstructed surface with the (110) orientation after annealing at 200~$^\circ$C under an oxygen partial pressure of $1\times10^{-5}$~Torr. However, this reconstructed surface was observed at temperatures both above and below the MIT and XPS confirms a change in the oxygen content of the film surface after the anneal (see Supplementary Figure 3). Furthermore, after irradiating a sample region with the electron beam for several minutes, the fractional LEED spots get weaker and the pattern reverts to the original ($1\times1$) pattern. A clear change in work function is also observed with LEEM after this extended electron beam exposure. This behavior implies the observed ($2\times2$) reconstruction is related to oxygen adsorption/desorption on the film surface, instead of the bulk structural transition. This is further supported by our DFT calculations which demonstrate larger variation in work function as oxygen is adsorbed/desorbed on the rutile $(110)_R$ surfaces as opposed to transitioning from a rutile to monoclinic $(011)_M$ surface (see Supplementary Table~I).
	
	Of note, the (100) and (001) surfaces do not have this behavior when using an oxygen partial pressure of $1\times10^{-5}$~Torr, indicating this reconstruction is specific to the (110) surface. This may be due to the more polar nature of the (110) film compared to both (100) and (001). Our DFT calculations demonstrate that cleaving the bulk crystal structure at different facets theoretically produces surface configurations with varying oxygen content.
	
	We used DFT to investigate oxygen vacancy/adsorption on the \ce{VO2}(110) surface by removing V (adsorption) or O (vacancy) from \ce{VO2} slabs. In this context, $\Gamma$ represents the excess number of O ions at each surface of the slab. Only the (1$\times$1) rutile slab shown in Figure \ref{VO2_110_damage}c was stable under a stoichiometric configuration for the \ce{VO2}(110) surface, while a (2$\times$2) reconstruction was obtained for an oxygen rich surface ($\Gamma>1$). Slabs with an intermediate oxygen excess/deficiency ($\Gamma=\pm1$) were predicted to have a ($\sqrt{2}\times\sqrt{2}$)-like surface symmetry. 
	
	More importantly, both the $\Gamma=1$ and (2$\times$2) reconstructed $\Gamma=3$ oxygen rich slabs were predicted to possess vanadyl bonds on the surface (see Supplementary Figure 4). These surface stabilizing vanadyl terminations will effectively push the orbital occupation of the surface vanadium away from the Mott criterion (e.g., $d^n \to 0$), similar to Ti doping in the bulk,\cite{Quackenbush2017} which should result in a complete loss of Mott MIT physics.  
	
	As shown in Figure \ref{VO2_110_damage}d, these oxidized surfaces are more likely to be stable than the reduced surfaces for the $\Delta\mu_O$ range corresponding to where bulk \ce{VO2} is thermodynamically stable with respect to \ce{V2O5} (-2.26 to -1.20~eV).\cite{Ong2013} This is true for all film orientations investigated in this study (see Supplementary Figure 5), explaining why the bulk transition of the epitaxial \ce{VO2} films could not be observed at any of the surface orientations investigated in this study.
	
	\section{Conclusions}
	
	The inherent vanadyl forming behavior of \ce{VO2} surfaces must be considered when developing future, ultra-thin, vanadium oxide devices, as it implies some minimum film thickness greater than a mono-layer would be required for switching in real-world devices. These results, combined with our previous work showing switching in \ce{VO2} as thin as $\sim$1 nm,\cite{Quackenbush2015} narrows the potential minimum thickness to between 1 to 3 unit cells. Furthermore, these results also have important implications for electrochemical applications of crystalline \ce{VO2}, where oxygen redox active surfaces may be desired.
	
	Surface reconstructions of \ce{VO2} are driven primarily by the lowering of surface energy through exposure of preferential planes and the formation (or removal) of vanadyl species. As such, they can form and evolve independently of the underlying bulk phase of the material. Even for the most correlated (Mott-like) strain orientation investigated in this study, \ce{VO2}/\ce{TiO2}(110),\cite{MukherjeePRB2016} we do not observe any evidence of the 'bulk' structural transition at the very surface. As a result, extreme caution must be taken when employing surface sensitive spectromicroscopy techniques to investigate MIT materials like \ce{VO2}, as the independent behavior of the surface could easily be misinterpreted as a decoupling or an appearance of exotic phases.

	\section{Methods}
	
	A set of high quality, epitaxial \ce{VO2} thin films were grown on rutile \ce{TiO2}(001), (100), and (110) single crystal substrates (Crystec GmbH, Germany) by reactive MBE in a Veeco GEN10 system at Cornell University. Substrates were first prepared by etching and annealing to have clean and well-defined step and terrace microstructured surfaces.\cite{Yamamoto2005} Vanadium and distilled ozone were co-deposited onto the substrate held at 250~$^\circ$C under a distilled ozone background pressure of 1.0$\times$10$^{-6}$~Torr. Following deposition of the desired film thickness, the temperature of the sample was rapidly ramped to 350~$^\circ$C, then immediately cooled to below 100~$^\circ$C under the same background pressure of distilled ozone to achieve an improved film smoothness and a more abrupt MIT. Sample thicknesses ranged from 1 to 10~nm for each orientation, as determined by monitoring the oscillations in reflection high energy electron diffraction (RHEED). Further details on the growth methodology have been previously described in the literature.\cite{Tashman2014,Paik2015,Quackenbush2016M2}
	
	The quality and MIT switching behavior of the films were investigated after growth using x-ray photoelectron and absorption spectroscopy before and after LEED/LEEM surface studies in order to ensure the film epilayer was not significantly altered by the employed surface preparation processes or by the measurements themselves. All measurements were performed under an ultra-high vacuum base pressure of $4\times10^{-10}$~Torr or better. A washing procedure was used to clean atmospheric contamination and over oxidation from the sample surfaces\cite{Surnev2003,Goering1997,Quackenbush2017} before performing any surface sensitive measurements (see Supplementary Figure 2 and the corresponding discussion). 
	
	Soft x-ray photoelectron spectroscopy (XPS) was performed at the Analytical and Diagnostics Laboratory (ADL) at Binghamton University using a laboratory-based monochromated PHI 5000 Versaprobe system. The spectra were measured using a monochromated Al K$\alpha$ x-ray source and a pass energy of 23.5 eV, corresponding to an instrumental resolution of 0.51 eV as determined from analyzing the Fermi edge and Au 4f$_{7/2}$ peak of Au foil references.
	
	Hard x-ray photoelectron spectroscopy (HAXPES) and x-ray aborption spectroscopy (XAS) were performed at the I09 beamline of the Diamond Light Source (DLS) in Oxfordshire, UK. HAXPES spectra were measured at a 5.934 keV photon energy using a VG Scienta EW4000 high-energy electron-energy analyzer with a $\pm$ 30$^\circ$ acceptance angle. The HAXPES photon beam was monochromated using a Si(111) double-crystal monochromator followed by a channel cut Si(004) crystal, resulting in an energy resolution of $<$ 250 meV.\cite{DiamondMono,WangohAPL2016,Quackenbush2016M2} Binding energy was calibrated using the Au 4f$_{7/2}$ peak and Fermi edge position of clean Au foil. 
	
	LEEM images were collected using Elmitec aberration-corrected low energy electron microscope (AC-LEEM) equipped with an energy analyzer. LEED measurements were performed using the LEEM optics, along with a microchannel plate detector and 1.5~$\mu$m selected-area aperture. Unlike traditional LEED, the positions of the diffraction spots on the detector do not depend on beam energy using this experimental geometry. This allows for the straightforward merging/averaging of multiple LEED images taken at different beam energies, which circumvents structure factor-related extinctions that occur at certain energies thus revealing the complete diffraction pattern.  Corresponding \ce{TiO2} (1$\times$1) reconstructed surfaces for each orientation were prepared and used for calibration of the \ce{VO2} LEED patterns.
	
	Scanning transmission electron microscopy (STEM) data were recorded from mechanically polished cross-sectional \ce{VO2/TiO2} specimens in the 100 keV NION UltraSTEM, a 5th order aberration corrected microscope optimized for EELS spectroscopic imaging with a probe size of $\sim$1 \angstrom and a beam current of $\sim$150 pA.
	
	All slab and bulk DFT calculations for the R and M1 phases of \ce{VO2} were performed using the Vienna Ab initio Simulation Package (VASP) within the projector augmented wave (PAW) approach.\cite{Blochl1994, Kohn1965, DFT4, Blochl1994} The exchange-correlation effects were modeled using the Perdew-Berke-Ernzerhof (PBE) generalized gradient approximation (GGA) functional.\cite{Perdew1996} We also applied the Hubbard U parameter with U=3.25 eV to all calculations in accordance with Wang et al.\cite{Wang2006h,Jain2011} as unlike GGA, GGA+U correctly demonstrates that the M1 phase is more stable than the R phase. 
	
	All calculations were performed with a plane wave cutoff of 500 eV. The pseudopotentials used are similar to those used in the Materials Project.\cite{Jain2013} The energies and atomic forces of all calculations were converged within $10^{-4}$ eV and 0.02 eV$\angstrom^{-1}$ respectively. Slabs were generated using the method described by Sun et al.\cite{Sun2013a} $\Gamma$-centered k-point meshes of $\frac{\bf{a}}{50}\times\frac{\bf{b}}{50}\times1$, where $\bf{a}$ and $\bf{b}$ are the the lattice parameters of the slab cell, were used for the slab calculations with non-integer values rounded up to the nearest integer. All input generation, surface energy analysis and Wulff shape generation was performed with the Python Materials Genomics (pymatgen) materials analysis library.\cite{Ong2013, Tran2016a}
	
	Like with previous DFT studies of \ce{VO2}\cite{Mellan2012, Eyert2011, Wickramaratne2019}, we omitted spin-polarization when comparing the surface energies (Table~\ref{VO2_TEM_LEEM}c) and work functions (Supplementary Table~I) of the monoclinic and rutile phases. However, both ferromagnetic and non spin-polarized surfaces were considered when investigating the non-stoichiometric surface energies of the rutile phase. While both magnetic configurations yield the same stable terminations, the non spin-polarized non-stoichiometric surface energies of the rutile phase yield unphysical (negative) surface energies under a high oxygen chemical potential  (see Supplementary Figure~5). As such, the ferromagnetic surface energies are shown in Figure~\ref{VO2_110_damage}d instead. For a further explanation of magnetic considerations in DFT, the reader is directed to the Supplementary Information.
	
	The surface energies used to determine the relative stability of all slab systems were calculated using the surface grand potential given by:
	\begin{equation}
	\label{eq:surface_energy}
	\gamma = \frac{1}{2A}[E^{slab}-\sum_{i}N_i\mu_i]
	\end{equation}
	where $E^{slab}$ is the total energy of the slab, $i$ is a unique species, $\mu_{i}$ is the chemical potential of species $i$ and $N_{i}$ is the total number of $i$ atoms in the slab, A is the surface area and the factor of 2 accounts for the slab model containing two symmetrically equivalent surfaces.\cite{Rogal2007} Further explanation and details into surface thermodynamics and slab generation can be found in the Supplementary Information.
	
	\section{References}
	
	
	%

	\section{Acknowledgments}
	
	We thank Marc Sauerbrey for his support in the initial phase of the project and Jens Falta for stimulating discussions. This work was supported by the Air Force Office of Scientific Research Multi-Disciplinary University Research Initiative (MURI) entitled, Cross-disciplinary Electronic-ionic Research Enabling Biologically Realistic Autonomous Learning (CEREBRAL), under award number FA9550-18-1-0024. We acknowledge Diamond Light Source for time on Beamline I09 under Proposals SI25355 and SI13812 for XAS measurements. This research used resources of the Center for Functional Nanomaterials, which is U.S. Department of Energy (DOE) Office of Science facility at Brookhaven National Laboratory, under Contract No. DE-SC0012704. We acknowledge support from the National Science Foundation [Platform for the Accelerated Realization, Analysis, and Discovery of Interface Materials (PARADIM)] under Cooperative Agreement No. DMR-1539918. Work was performed in part at the Cornell NanoScale Facility, a member of the National Nanotechnology Coordinated Infrastructure (NNCI), which is supported by the National Science Foundation (Grant No. ECCS-1542081). Work by R.T and S.P.O was supported by Quantum Materials for Energy Efficient Neuromorphic Computing (Q-MEEN-C), an Energy Frontier Research Center funded by DOE, Office of Science, BES under Award \# DE-SC0019273. J-OK and JIF gratefully acknowledge support by the Deutsche Forschungsgemeinschaft (DFG) under project nos. 362536548 and 408002857.

	\section{Author contributions}
	The contributions of authors are as follows. LFJP and NFQ conceived of the study. MJW and NFQ performed the bulk of the data analysis, generated the figures, and wrote the manuscript. MJW and NFQ performed the temperature dependent HAXPES/XAS. JTS, MJW, and NFQ performed the LEED/LEEM/UPS measurements. J-OK and JIF aided with the structural analysis and interpretation of the LEED results. JIF performed initial \ce{VO2} surface studies and helped develop the \ce{VO2} surface cleaning recipe. RT and SPO performed the surface stability calculations and corresponding DFT, as well as aided in preparing the manuscript for submission. CS and T-LL aided with the synchrotron experiments at the Diamond Light Source. MH and DAM performed the TEM imaging. HP and DGS are responsible for the sample fabrication and initial characterization of sample quality.
	
	\section{Additional information}
	Supplementary information is available in the online version of the paper. Reprints and permissions information is available online at www.nature.com/reprints.
	Correspondence and requests for materials should be addressed to 
	Louis F J Piper, lpiper@binghamton.edu
	Shyue Ping Ong, ongsp@eng.ucsd.edu
	
	\section{Competing financial interests}
	The authors declare no competing financial interests.

	\makeatletter
	\close@column@grid
	\clearpage
	\twocolumngrid
	\makeatother
	\newpage
	
\end{document}